\documentclass[12pt]{article}    
\usepackage{latexsym} 
\usepackage{amssymb}  
\usepackage{amsfonts} 
\usepackage{epsf}

\newcommand{\al}{\alpha}
\newcommand{\be}{\beta}
\newcommand{\ga}{\gamma}

\newcommand{\de}{\delta}
\newcommand{\De}{\Delta}
\newcommand{\ep}{\varepsilon}
\newcommand{\eps}{\epsilon}

\newcommand{\ka}{\kappa}

\newcommand{\La}{\Lambda}

\newcommand{\si}{\sigma}

\renewcommand{\th}{\theta}   

\newcommand{\<}{\langle} 
\renewcommand{\>}{\rangle} 

\newcommand{\txt}{\textstyle}
\newcommand{\dsp}{\displaystyle}

\newcommand{\ad}{\dagger}
\newcommand\eqn[1]{(\ref{#1})}      
\newcommand{\e}{ {\rm e} }
\newcommand{\beq}{\begin{equation}}
\newcommand{\eeq}{\end{equation}}
\newcommand{\ba}{\begin{array}}
\newcommand{\bea}{\begin{eqnarray}}
\newcommand{\ea}{\end{array}}
\newcommand{\eea}{\end{eqnarray}}

\newcommand\comment[1]{ \hbox{[{\it Comment suppressed here.}\/]} }
\newcommand\hide[1]{}

\newcommand{\skipover}[1]{}

\newcommand{\half} {{\txt {1\over 2}}}
\newcommand{\third}{{\txt {1\over 3}}}
\newcommand{\quarter}{{\txt {1\over 4}}}

\newcommand{\eighth}{{\txt {1\over 8}}}

\def\phm{\phantom{-}}

\newcommand{\cc}{ {\rm c.c.} }

\newcommand{\MeV}{{\rm MeV}}
\newcommand{\GeV}{{\rm GeV}}

\def\appendix{\par                              
    \setcounter{section}{0}                     
    \setcounter{subsection}{0}
    \renewcommand{\theequation}{\Alph{section}.\arabic{equation}}
    \renewcommand{\thesection}{Appendix \Alph{section}
                \setcounter{equation}{0}  } 
}

\catcode`\@=11

\def\applabel#1{\@bsphack
  \protected@write\@auxout{}%
         {\string\newlabel{#1}{{\Alph{section}}{\thepage}}}%
  \@esphack}

\def\section{
\setcounter{equation}{0}        
\@startsection {section}{1}{\z@}{-3.5ex plus -1ex minus 
 -.2ex}{2.3ex plus .2ex}{\large\bf}}
\renewcommand{\theequation}{\arabic{section}.\arabic{equation}}

\def\subsection{\@startsection{subsection}{2}{\z@}{-3.25ex plus -1ex minus 
 -.2ex}{1.5ex plus .2ex}{\normalsize\bf}}

\def\subsubsection{\@startsection{subsubsection}{3}{\z@}{-3.25ex plus
 -1ex minus -.2ex}{1.5ex plus .2ex}{\normalsize}}

\newsavebox{\eqlabel}

\makeatletter  
\newlength{\numblen}
\newsavebox{\eqnumb}
\def\@eqnnum{\savebox{\eqnumb}{\rm (\theequation)}%
\settowidth{\numblen}{\usebox{\eqnumb}}%
\makebox[\numblen][l]{\usebox{\eqnumb}~~~\usebox{\eqlabel}}}
\makeatother   

\newenvironment{equationwithlabel}[1]{ %
  \begin{equation}\label{#1} }{\end{equation}} 
\newcommand{\beql}[1]{\begin{equationwithlabel}{#1}}
\newcommand{\eeql}{\end{equationwithlabel}}

\newcommand{\psib}{{\bar\psi}}
\newcommand{\psid}{{\psi^\ad}}
\newcommand{\psiL}{{\psi_L}}

\newcommand{\psiLd}{{\psi_L^\ad}}
\newcommand{\psiRd}{{\psi_R^\ad}}

\newcommand{\da}{{\dot a}}

\newcommand{\dc}{{\dot c}}

\newcommand{\ab}{{\bar a}}
\newcommand{\bb}{{\bar b}}
\newcommand{\aL}{{a_L^{\phantom\ad}}}
\newcommand{\aR}{{a_R^{\phantom\ad}}}
\newcommand{\bL}{{b_L^{\phantom\ad}}}
\newcommand{\bR}{{b_R^{\phantom\ad}}}
\newcommand{\aLd}{{a_L^\ad}}
\newcommand{\aRd}{{a_R^\ad}}
\newcommand{\bLd}{{b_L^\ad}}
\newcommand{\bRd}{{b_R^\ad}}
\newcommand{\yL}{{y_L^{\phantom\ad}}}

\newcommand{\zL}{{z_L^{\phantom\ad}}}

\newcommand{\yLd}{{y_L^\ad}}

\newcommand{\zLd}{{z_L^\ad}}

\newcommand{\bp}{{\bf p}}
\newcommand{\bk}{{\bf k}}

\newcommand{\xiy}{\xi^y}

\newcommand{\xiz}{\xi^z}

\newcommand{\FF}{\mathfrak{F}} 
\newcommand{\Z}{\mathbb{Z}} 

\begin{document}

\title{ {\bf Color-Flavor Locking and\\ 
Chiral Symmetry Breaking \\
in High Density QCD} }

\newcommand{\ns}{\normalsize}

\author{
Mark Alford${}^{(a)}$, Krishna Rajagopal${}^{(b)}$, 
Frank Wilczek${}^{(a)}$ \\[0.5ex]
{\normalsize 
 ${}^{(a)}$ School of Natural Sciences, Institute for Advanced Study,
Princeton, NJ 08540} \\[0.5ex]
{\normalsize ${}^{(b)}$ Center for Theoretical Physics,}\\
{\normalsize Massachusetts Institute of Technology, Cambridge, MA 02139 }
}

\newcommand{\preprintno}{\normalsize IAS-SNS-98/29, MIT-CTP-2731,
hep-ph/9804403
}

\date{\preprintno}

\begin{titlepage}
\maketitle
\def\thepage{}          

\begin{abstract}
We propose a symmetry breaking scheme for QCD with three massless
quarks at high baryon density wherein the color and flavor SU(3)$_{\rm
color}\times$SU(3)$_L\times$SU(3)$_R$ symmetries are broken down to
the diagonal subgroup SU(3)$_{{\rm color}+L+R}$ by the formation of a
condensate of quark Cooper pairs.  We discuss general properties that
follow from this hypothesis, including the existence of gaps for quark
and gluon excitations, the existence of Nambu-Goldstone bosons which
are excitations of the diquark condensate, and the existence of a
modified electromagnetic gauge interaction which is unbroken and which
assigns integral charge to the elementary excitations.  We present
mean-field results for a Hamiltonian in which the interaction between
quarks is modelled by that induced by single-gluon exchange.  We find
gaps of order 10-100 MeV for plausible values of the coupling.  We
discuss the effects of nonzero temperature, nonzero quark masses and
instanton-induced interactions on our results.

\end{abstract}

\end{titlepage}

\renewcommand{\thepage}{\arabic{page}}


\section{Introduction}
\label{sec:int}

The behavior of matter at high quark density is interesting in itself 
and is relevant to phenomena in
the early universe, in neutron stars, and in heavy-ion
collisions.  Unfortunately, the presence of a chemical potential makes
lattice calculations impractical, so our understanding of high density
quark matter is still rudimentary.  
Even qualitative questions, such as 
the symmetry of the ground state, are unsettled.  

In an earlier paper \cite{ourselves} we considered an idealization of
QCD, supposing there to exist just two species of massless quarks.  We
argued that at high density there was a tendency toward spontaneous
breaking of color (color superconductivity).  Specifically, we
analyzed a model Hamiltonian with the correct symmetry structure,
abstracted from the instanton vertex, 
and found an instability 
toward the formation of Cooper pairs of quarks
leading to a sizable condensate of the form 
\beql{oldCondensate} 
\langle q_i^{\alpha} C \gamma^5 q_j^{\beta}\rangle 
~\propto ~\epsilon_{ij} \epsilon^{\alpha\beta3}~,
\eeql
where the Latin indices signify flavors and the Greek indices signify
colors.  
The choice of ``3'' for the preferred direction in color space
is, of course, conventional.
In the presence of the condensate (\ref{oldCondensate}) the color symmetry
is broken down to SU(2), while chiral SU(2)$_L\times$SU(2)$_R$ is left
unbroken.  Other authors have reached similar conclusions
\cite{bailin,instantonliquid}, and color breaking condensates 
have also been studied in a single-flavor model \cite{iwaiwa}.  
There is an obvious asymmetry in the {\em ansatz\/}
\eqn{oldCondensate} between the two ``active'' colors that
participate in the condensation, and 
the third, passive one.\footnote{We did find a very much weaker tendency
toward condensation of the quarks of the third color
in an isoscalar axial vector channel \cite{ourselves}. The third color
quarks, therefore,  
need not be completely passive but this 
phenomenon makes the asymmetry
between their behavior and that of the first two colors even more
pronounced.}  
Some
such asymmetry is inevitable in a world with two light quarks due to
the mismatch between the number of colors and the number of flavors.

If the strange quark were heavy relative to fundamental QCD scales,
the idealization involved in assuming two massless flavors would be
entirely innocuous. In reality, this is not the case as
we now explain.
We are interested in densities above the transition
at which the ordinary chiral condensate $\langle \bar q q\rangle$
vanishes (or, in the case of $\langle \bar s s \rangle$, is
greatly reduced.) This means that the strange quark
mass is approximately equal to
its Lagrangian mass $m_s$, which is of order 100 MeV.
Color superconductivity
involves quarks near the Fermi surface.  As we are interested
in chemical potentials which are large compared to $m_s$,
strange quarks cannot be neglected.
Furthermore,
one can 
expect condensates at the Fermi surface 
to be essentially unaffected by  
the presence of quark masses that are small compared
to the chemical potential, and this has been demonstrated
explicitly for the condensate
(\ref{oldCondensate}) in the two flavor model \cite{withjurgen}.
As we are interested in chemical potentials which are large compared to
$m_s$, taking $m_s=0$ is a reasonable starting point
for the physics of interest to us here.

Thus on both formal and physical grounds it is of considerable
interest to consider an alternative idealization of QCD, supposing
there to exist three species of massless quarks.  
We shall argue that upon making this idealization,  
there is a symmetry breaking pattern,
generalizing \eqn{oldCondensate}, which has several interesting and
attractive features including
the participation of quarks of all
flavors and colors and the breaking
of both color and flavor symmetries,
and present model calculations which indicate
that this pattern is likely to be energetically favorable.


\section{Proposed Ordering} 
\label{sec:ordering}

\def\SU3{\rm SU(3)}
\def\U1{\rm U(1)}
The symmetry of QCD with three massless quarks is $\SU3_{\rm
color}\times\SU3_L\times\SU3_R\times \U1_B$.  The SU(3) of
color is a local gauge symmetry, while the chiral flavor SU(3)
symmetries are global, and the final factor is baryon number.  Our
proposal is that at high density this symmetry breaks to the diagonal
SU(3) subgroup of the first three factors -- a purely global symmetry.

A condensate invariant under the diagonal $\SU3$ is
\beql{twoCondensates}
\langle q_{Lia}^\alpha\, q_{Ljb}^{\beta}\,\epsilon^{ab} \rangle = 
-\langle q_{Ri}^{\alpha\dot a} \,q_{Rj}^{\beta\dot b}\,\epsilon_{\dot a
\dot b} \rangle 
= \kappa_1 \delta^\alpha_i \delta^\beta_j + \kappa_2 \delta^\alpha_j 
\delta^\beta_i~,
\eeql
where we have written the condensate using two component Weyl 
spinors. The explicit Dirac indices make the $L$'s and $R$'s 
superfluous here, but we will often drop the indices.
The mixed Kronecker $\delta$ matrices are
invariant under matched 
vectorial color/flavor
rotations, leaving only the diagonal $\SU3$ unbroken.
That is, the $LL$ condensate ``locks'' $\SU3_L$ rotations
to color rotations, while the $RR$ condensate locks $\SU3_R$ 
rotations to color rotations.  As a result, the only remaining
symmetry is the global symmetry $\SU3_{{\rm color}+L+R}$
under which one makes simultaneous rotations in color and
in vectorial flavor. In particular, the gauged color symmetries
and the global axial flavor symmetries are spontaneously broken.

The condensates in (\ref{twoCondensates}) can be written
using Dirac spinors as 
$\langle q^\alpha_i C\gamma^5  q^\beta_j \rangle$, and
therefore constitute a Lorentz scalar.
Note that it might be possible for the $LL$ and $RR$
condensates in (\ref{twoCondensates}) to differ other than by a sign.
This would violate parity.  As discussed further below, we have
seen no sign of 
this phenomenon in the model Hamiltonians which we consider.
Note also that when $\ka_1 = -\ka_2$,  the condensate can be rewritten
as $\langle q_i^{\alpha} C\gamma^5  q_j^{\beta}\rangle 
~\propto ~\epsilon_{ijI} \epsilon^{\alpha\beta I}$, which is
a natural generalization of the two-flavor ordering \eqn{oldCondensate}.
This would mean that the Cooper pair wave functions would
be antisymmetric under exchange of either color or flavor.
We find, however, that solutions of the 
coupled gap equations 
for $\ka_1$ and $\ka_2$ 
do not have solutions with $\ka_1 = -\ka_2$. The Cooper pairs
in the condensate (\ref{twoCondensates}) are symmetric under
simultaneous exchange of color and flavor, but are not antisymmetric
under either color or flavor exchange. 

The ordering (\ref{twoCondensates}) is not the only possibility.
One could first of all try a state which is antisymmetric
under simultaneous exchange of color and flavor, but which
has spin one.  Based on the results of Ref. \cite{ourselves},
this is very unlikely to compete with (\ref{twoCondensates}),
because not all momenta at the Fermi surface can participate equally.
A much more plausible possibility is that, when one allows $m_s$
to differ from the up and down quark masses, one will have a
less symmetric condensate in which the strange quarks are 
distinguished from the up and down quarks.  Although we have
argued above that the strange quark mass in nature is light enough
that $m_s=0$ is a reasonable starting point for estimating
the magnitude of the superconductor gap, it is certainly
the case that once $m_s$ is set to its physical value, the 
symmetry of the condensate will be reduced.\footnote{As
an exercise, we have added a four-fermion interaction involving
up and down quarks only, modelled on the 't~Hooft vertex in
the two flavor theory.   This mocks up (some of) the effects
of the true six-fermion 't~Hooft vertex at 
nonzero $m_s$, as long as the coupling constant is taken
to be proportional to $m_s$.  We find five independent gap parameters
instead of two. That is, the condensates are less symmetric, as 
expected.  As the coupling of the instanton-like 
four-fermion interaction
is increased, 
the condensate changes smoothly from (\ref{twoCondensates})
to (\ref{oldCondensate}).  A full treatment using the 
six-fermion interaction remains to be done, 
but we expect this crossover to occur for strange
quark masses of order $\mu$.}

To forestall a possible difficulty, let us remark that the formal
violation of flavor and baryon number by our condensate does not imply
the possibility of large genuine violation of flavor or baryon number
in the context of a heavy ion collision or a neutron star, any more
than ordinary superconductivity (with Cooper pairs) involves violation
of lepton number.  The point is that the condensates occur only in a
finite volume, and the conservation equation can be integrated over a
surface completely surrounding and avoiding this volume, allowing one
to track the conserved quantities.  There is, however, the possibility
of easy transport of quantum numbers into or out of the affected
volume, and thus of large dynamical fluctuations in the normally
conserved quantities in response to small perturbations.  These are
the typical manifestations of superfluidity.

In the following sections we shall present quantitative calculations,
based on an interaction Hamiltonian modelling single-gluon exchange,
which indicate that symmetry breaking of this form, with substantial
condensates, becomes energetically favorable at any density
which is high enough that the ordinary $\bar q q$ condensate
vanishes.  Before
describing these calculations in detail, however, we would like to
make a number of general qualitative remarks.

{\it Motivation}: As already discussed, our proposal here for three
flavors is a natural generalization of what we and others have already
found to be favorable for two flavors.  Since it retains a high degree
of symmetry -- the residual SU(3), as well as space-time rotation
symmetry -- different flavor sectors and different parts of the Fermi
surfaces all make coherent contributions, thus taking maximal
advantage of the attractive channel.  The proposed ordering bears a
strong resemblance to the B phase of superfluid ${}^3$He,
which is the ordered state of liquid ${}^3$He 
favored at low temperatures.
In that phase, atoms form Cooper pairs such that
vectors associated with orbital and nuclear spin degrees of freedom
are correlated, breaking an SO(3)$\times$SO(3) symmetry to the
diagonal subgroup.

{\it Broken chiral symmetry}: Chiral symmetry is spontaneously broken
by a new mechanism: locking of the flavor 
rotations to color.

{\it Gap}: The condensation produces a gap for quarks of all
three colors and all three flavors, for
all points on the Fermi surface.  Thus there are no residual low
energy single-particle excitations.  The color superconductivity
is ``complete''.  This has immediate consequences in neutron stars.
The rate of neutrino emission from
matter in this phase is exponentially suppressed for temperatures
less than of order the gap.  In the two flavor theory, this 
rate was only reduced by a factor $2/3$, unless the third color
quarks were also able to condense.  

{\it Higgs phenomenon and Nambu-Goldstone bosons}: The symmetry has
been reduced by 17 generators.  Of these, eight were local generators.
The corresponding quanta -- seven gluons and
one linear combination of the eighth gluon and the photon
of electromagnetism -- all acquire mass
according to the Higgs phenomenon.  
Eight other generators correspond to the
broken chiral symmetry.  They generate massless collective
excitations, the Nambu-Goldstone bosons.  The broken symmetry
generators are given by the axial charges, just as in the standard
discussion of chiral symmetry breaking in QCD at zero density.  
Goldstone's theorem entails the existence of massless bosons
with the quantum numbers of the broken symmetry generators,
again as at zero density.  They must be states of 
negative parity and zero spin.
Finally, there is an additional scalar 
Nambu-Goldstone particle associated
with spontaneous breakdown of baryon number symmetry.  
Of course the Nambu-Goldstone bosons associated with chiral symmetry
breaking form an octet under the unbroken SU(3), while the
Nambu-Goldstone boson associated with baryon number violation is a
singlet.

The octet Nambu-Goldstone bosons
are created by acting with an 
axial charge operator on a diquark condensate which
spontaneously breaks $\U1_B$.  This means that they
do not have well-defined baryon number.  This
is evident once one realizes that a propagating $\bar q \gamma_5 q$
oscillation
can become a propagating $\bar q C \bar q$ (or $q C q$) oscillation
by an interaction with the $\langle \bar q C\gamma_5 \bar q\rangle$ 
(or $\langle q C\gamma_5 q\rangle$) condensate 
in which a Nambu-Goldstone boson
associated with the breaking of $\U1_B$ is excited.  
In a lattice simulation, the octet Nambu Goldstone modes
could be created by inserting $qCq$, $\bar q C\bar q$ or $\bar q \gamma_5 q$
operators.  

One might be concerned with the use of diquark operators, which are
not gauge invariant, as interpolating fields for the Nambu-Goldstone bosons.
We do not think this is a serious difficulty; but in any case, we could 
alternatively use operators of the form
$\bar q \gamma_5 q$ and $(qqq)^2$ for the 
chiral symmetry octet and baryon number singlet, respectively.

{\it Residual $\Z_2\times \Z_2$, axial baryon number, and parity}: 

Our model Hamiltonian is symmetric under $\U1_V \times \U1_A$. 
Our condensate breaks this down to a $\Z^L_2\times \Z^R_2$
that changes the sign of the left/right- handed quark fields. In
real three-flavor QCD, $\U1_A$ is anomalous, and is broken to $\Z_6$ by
instantons. The condensate would then break $\U1_V \times \Z_6$ down to
the common $\Z_2$ which flips the sign of all quark fields.

Because our Hamiltonian does not explicitly violate $\U1_A$, we
cannot use it to  
infer the relative phase of the $LL$ and $RR$ 
condensates.  In our previous work on the two-flavor case we
used a model Hamiltonian abstracted from the instanton, that did
violate the anomalous symmetries.  There we found that the
parity-conserving choice of relative phases was favored.  This
interaction will still be present, of course, and we expect that the
parity-conserving phase will still be favored dynamically in real QCD.
Hence, we choose the minus sign in (\ref{twoCondensates}) which
makes the condensate a Lorentz scalar.
Upon making this choice, the Nambu-Goldstone bosons associated with axial
flavor symmetry breaking will be pseudoscalar, and the Nambu-Goldstone
boson associated with baryon number violation will be scalar.
In the 3-flavor theory, the 't~Hooft vertex has six fermion
legs. This means that it is irrelevant, in the sense that its
coefficient is reduced as modes are integrated out and only
those closer and closer to the Fermi surface are kept. 
Thus in practice
its effects might become quite small at high density.  
In that case there will
be an additional light pseudoscalar, associated with axial baryon number,
which is a singlet under the unbroken SU(3) and whose mass does not
vanish in the chiral limit.


{\it Effect of quark masses}: Quark mass terms are present in the real
world.  They lift the masses of the pseudoscalar octet of
Nambu-Goldstone bosons.  One can consider this effect along the lines
set out by Gell-Mann, Oakes and Renner for conventional chiral
symmetry breaking.  Explicit breaking of chiral symmetry
occurs through inclusion of symmetry breaking operators in the
effective Lagrangian.  The
addition of an $m_q \bar q q$ operator 
to the Lagrangian yields $m_{NG}^2 \propto m_q$ if 
$\langle \bar q q \rangle$ is nonzero.  If $\Z_2^L$ were a valid symmetry,
this contribution would vanish.  Then one must
go to higher order and
consider the operator proportional to $m_q^2 (\bar q q)^2$.
The expectation value of this operator is nonzero, as it
can be written as a product of $\langle q C\gamma_5 q\rangle$ and 
$\langle\bar q C \gamma_5 \bar q\rangle$ 
condensates, and so we obtain a contribution to $m_{NG}^2 \propto m_q^2$,
which is likely small (but see below).

The scalar Nambu-Goldstone boson associated with baryon number
violation of course remains strictly massless, since baryon
number symmetry is not 
violated by quark mass terms.

{\it Effects of instanton-induced interactions}:
We have already seen that these interactions, although small,
are needed to fix the relative sign of the condensates with
differing helicities and to give a (small) mass to the
$\eta'$-like boson.  In addition, by breaking
$\Z_2^L$, they induce a further qualitative
modification.  The 't~Hooft vertex 
can connect an incident $\bar q_L$  and $q_R$ to the product
of the $\langle \bar q_R \bar q_R \rangle$ and $\langle q_L q_L \rangle$
condensates.  This means that in the presence of both the condensate
(\ref{twoCondensates}) and the instanton-induced interaction,
one can  have an ordinary $\langle \bar q_R q_L \rangle$ condensate.
This breaks no new symmetries: chiral symmetry is already broken.
As noted above, the 't~Hooft vertex is irrelevant
at the Fermi surface. 
The gap equation 
for the ordinary chiral condensate will not have a log 
divergence, and so the induced $\langle \bar q_R q_L \rangle$ 
condensate can be either nonzero or zero.  If it is nonzero, one must
analyze this condensate coupled with the coexisting diquark condensates, 
for example using the
methods of Ref. \cite{withjurgen}.  We leave this
analysis to future work, but note here that this effect
introduces a contribution to $m_{NG}^2 \propto m_q$.


{\it Modified electromagnetism}: Though the standard electromagnetic
symmetry is violated by our condensate, as are all the color gauge
symmetries, there is a combination of electromagnetic and color
symmetry that is preserved.  This is possible because the
electromagnetic interaction is traceless and vectorial in flavor
SU(3).  Consider the gauged $\U1$ which is the sum of
electromagnetism, under which the charges of the quarks are
(2/3,-1/3,-1/3) depending on their flavor, and the color hypercharge
gauge symmetry under which the charges of the quarks are
(-2/3,1/3,1/3) depending on their color.  It is a simple matter to
check that the condensates (\ref{twoCondensates}) are invariant under
this rotation.  There is therefore a massless Abelian gauge boson,
corresponding to a modified photon.  In this superconductor,
therefore, although seven gluons and one linear combination of gluon
and photon get a mass and the corresponding non-Abelian fields display
the Meissner effect, there is a massless modified photon and a
corresponding magnetic field which can penetrate the matter.

Under this modified electromagnetism, the quark charges are compounded
{}from the (2/3, -1/3, -1/3) of their flavor and (-2/3, 1/3, 1/3) of
their color and thus are all integral, as are the charges of the
Nambu-Goldstone bosons.  Specifically, four of the Nambu-Goldstone
bosons have charges $\pm 1$, and the rest are neutral. These are just
the charges carried by the ordinary octet mesons under ordinary
electromagnetism.  Four of the massive vector bosons arising from the
color gluons via the Higgs mechanism have charges $\pm 1$, and the
rest are neutral.  The true Nambu-Goldstone boson associated with
spontaneous baryon number violation is neutral.

It is amusing that with quark masses turned on, all hadronic
excitations (single fermion excitations, massive gauge bosons, and
pseudo-Nambu-Goldstone bosons) with charges under the modified
electromagnetism acquire a gap.  Therefore, at zero temperature a
propagating modified photon with energy less than the lightest charged
mode cannot scatter, and the ultradense material is transparent.
Presumably some relatively small density of electrons will be needed
to ensure overall charge neutrality (since the strange quark is a bit
heavier than the others), and this metallic fraction will dominate the
low-energy electromagnetic response.

{\it Thermal properties}: 
For temperatures less
than the mass of the pseudo-Nambu-Goldstone bosons associated with
chiral symmetry breaking, 
and less than the gap, the specific heat is dominated
by the exact Nambu-Goldstone boson of broken
baryon number.  Charged excitations are exponentially suppressed.
This is quite unlike the physics
expected at high densities in the absence of a gap, where
one expects massless gluons and quasi-particle excitations
with arbitrarily low energies.  We find neither. 
At temperatures above the pseudo-Nambu-Goldstone mass, but below the gap,
the electromagnetic response will be
dominated by the pseudo-Nambu-Goldstone bosons.  As four of them
are charged under the modified $\U1$, they can emit
and scatter the modified photons.  At these temperatures,
the matter is no longer transparent.

Although in this paper we only present calculations at $T=0$, 
our proposal clearly 
raises interesting issues for the phase diagram, which
can be explored as has been done in the two-flavor 
theory\cite{withjurgen}.  
Here we will only make
a few simple remarks.  In the theory with three massless quarks, 
there will be a phase transition at some $\mu$-dependent temperature,
above which the diquark condensate vanishes 
and chiral symmetry and $\U1_B$ are restored. Even
with nonzero quark masses, this transition involves 
the restoration of the global $\U1_B$ symmetry, and 
cannot be an analytic crossover.
The existence of a broken $\U1$ in the high density phase
also ensures that it cannot be connected to the zero density phase without 
singularity,  despite the fact that both phases exhibit broken chiral  
symmetry.  Presumably  there is in fact a first-order transition
at low temperatures 
as a function of increasing chemical potential, with
a bag-model interpretation, similar to the
one we found for two flavors\cite{ourselves}.  
However now the phase interior to the bag
has {\it less\/} symmetry than the phase outside!  To join them, one will need
to quantize the collective coordinates of the broken symmetry generator for
baryon number.

We turn now to our model calculation of the two gaps $\kappa_1$ and 
$\kappa_2$
appearing in (\ref{twoCondensates}).

\section{Model Hamiltonian}
\label{sec:ham}

In our previous study of two-flavor diquark condensation, we used the
instanton vertex for our NJL model.
In the three flavor case, instantons are not
the dominant source of attractive interactions in the diquark
channel, since the instanton vertex now has 6 legs, and cannot
be saturated by diquark condensates.

However single-gluon exchange does provide an attraction between
quarks, and so we use an NJL model
containing a four-fermion interaction
with the color, flavor, and spinor structure of single-gluon exchange:
\beql{ham:H}
\ba{rcl}
H &=& \dsp\int d^3x\, \psib(x) (
\nabla\!\!\!\!/ - \mu\ga_0) \psi(x) + H_I, \\[2ex]
H_I &=& K \dsp\sum_{\mu, A}\int d^3x\, \FF\,
 \psib(x) \ga_\mu T^A \psi(x)\, \psib(x) \ga^\mu T^A \psi(x)\ ,
\ea
\eeql
where $T^A$ are the color $SU(3)$ generators.
In real QCD the interactions become weak at high momentum, so
we have included a schematic form factor $\FF$.
When we expand $H_I$ in momentum modes
we will make this factor explicit, via
a form factor $F(p)$ on each leg of the interaction vertex.
We will explore both smoothed-step and power-law
profiles for $F$:
\beql{ham:F}
F(p)= \dsp \left( 1 + \exp \Bigl[  {p-\La \over w} \Bigr] \right)^{-1} 
{\rm or}\quad 
F(p) = \Bigl( {\Lambda^2\over p^2+\La^2} \Bigr)^\nu,\qquad ;
\eeql

As noted above, a plausible scenario is that at high densities a
diquark condensate will form, breaking the color and flavor symmetries
down to their diagonal $SU(3)$ subgroup.  In Section~\ref{sec:SCgap}
we solve the mean-field gap equations for such a condensate, in order
to estimate the size of the gap parameters as a function of the
interaction strength $K$.  We derive the gap equations via the
Bogoliubov-Valatin approach, which is equivalent to the variational
method used in Ref. \cite{ourselves} but is perhaps simpler.  In order
to get some idea of the correct size of $K$, we perform a calculation
(Appendix A) of the interaction $H_I$ to the zero density chiral gap.
We present our results in Section~\ref{sec:results}.

\section{Color-Flavor Gap equations}
\label{sec:SCgap}

Single gluon exchange cannot convert left-handed massless
particles to right-handed, so we can rewrite the Hamiltonian in terms
of Weyl spinors, following only the left-handed particles from now
on. The computation for the right-handed particles would be identical.
(There are also terms in $H_I$ which couple left-
and right-handed quarks, but these do not contribute
to the gap equations for the $LL$ and $RR$ condensates of
(\ref{twoCondensates}).) 
Explicitly displaying color ($\al,\be\ldots$), flavor ($i,j\ldots$),
and spinor ($a,\da\ldots$) indices, and rewriting the color 
and flavor generators,
\beql{HI}
H_I = {2\over 3}K \int d^3x\, \FF \,
(3 \de^\al_\de \de^\ga_\be - \de^\al_\be \de^\ga_\de)
(\psid^i_{\al\da}\psid^j_{\ga\dc} \eps^{\da\dc}
 \psi^\be_{ib}\psi^\de_{jd} \eps^{bd}),
\eeql
where all fields are left-handed. We give our spinor conventions
in Appendix B.
Now make the mean-field ansatz 
\eqn{twoCondensates} for $|\psi\>$, the true ground state at
a given chemical potential:
\beql{Col:mf}
\ba{rcl}
\<\psi| \psid^i_{\al\da}\psid^j_{\ga\dc} \eps^{\da\dc} |\psi \>
 &=& 
\dsp {3\over 4 K} P^i_\al{}^j_\ga, \\[2ex]
P^i_\al{}^j_\ga &=&  
  \third(\De_8 + \eighth\De_1)\de^i_\al\de^j_\ga 
+ \eighth\De_1 \de^i_\ga \de^j_\al\ ,
\ea
\eeql
where the numerical factors have been chosen so that
$\De_1$ and $\De_8$, which parameterize $P$ and
which are linear combinations of $\kappa_1$ and $\kappa_2$, 
will turn out to be 
the two gaps (up to a form factor --- see the end of this section).
This condensate is invariant under the diagonal $SU(3)$
that simultaneously rotates color and flavor.
It therefore ``locks'' color and left-flavor rotations together
as described in Section~\ref{sec:ordering}.  
The interaction Hamiltonian becomes
\beq
\ba{rcl}
H_I &=& \dsp \half \int d^3x\, \FF\, Q_\be^i{}_\de^j \,
\psi^\be_{ib}\psi^\de_{jd} \eps^{bd} + \cc, \\[2ex]
Q_\be^i{}_\de^j &=& 
 \De_8 \de^i_\de\de^j_\be + \third(\De_1-\De_8) \de^i_\be \de^j_\de
\ea
\eeq
Replacing indices $i,\be$ with a single color-flavor index $\rho$,
we can simultaneously diagonalize the $9\times 9$ matrices 
$Q$ and $P$, and find that they have two eigenvalues,
\beq
\ba{rcl@{\qquad}rcl}
P_1 &=& \De_8 + \quarter\De_1,  &  P_2\cdots P_9 &=& \pm \eighth\De_1 \\[0.5ex]
Q_1 &=& \De_1,  &  Q_2\cdots Q_9 &=& \pm \De_8 \\
\ea
\eeq
That is, eight of the nine quarks in the theory have
a gap parameter given by $\Delta_8$, while the remaining
linear combination of the quarks has a gap parameter $\Delta_1$.
The Hamiltonian can be rewritten
in this color-flavor basis in terms of particle/hole creation/annihilation
operators $\ab_\rho,\bb_\rho$.
We also expand in momentum modes using \eqn{sp:modexp} and 
now explicitly include the
form factors $F(p)$ described in Section~\ref{sec:ham}. 
\beql{Col:H}
\ba{rcl}
H &=& \dsp\sum_{\rho,k>\mu} (k-\mu)\ab^\ad_\rho(\bk) \ab_\rho(\bk) 
+ \dsp\sum_{\rho,k<\mu} (\mu-k) \ab^\ad_\rho(\bk) \ab_\rho(\bk)
+ \dsp\sum_{\rho,\bk} (k+\mu) \bb^\ad_\rho(\bk) \bb_\rho(\bk) \\[2ex]
&+& \dsp \half \sum_{\rho,\bp}
F(p)^2 Q_\rho \e^{-i\phi(\bp)}\Bigl( 
   \ab_\rho(\bp) \ab_\rho(-\bp) 
+  \bb^\ad_\rho(\bp) \bb^\ad _\rho(-\bp)
\Bigr) + \cc,
\ea
\eeql
where the perturbative ground state, annihilated by $\ab_\rho$ and $\bb_\rho$
is the Fermi sea, with states up to $p_F=\mu$ occupied.
Finally, we change basis to creation/annihilation operators 
$y$ and $z$ for quasiparticles and quasiholes, 
\beql{Col:yz}
\ba{rcl}
y_\rho(\bk) &=& \cos(\th^y_\rho(\bk))\ab_\rho(\bk) + \sin(\th^y_\rho(\bk))
\exp(i\xiy_\rho(\bk)) \ab^\ad_\rho(-\bk) \\[1ex]
z_\rho(\bk) &=&  \cos(\th^z_\rho(\bk))\bb_\rho(\bk) + \sin(\th^z_\rho(\bk))
\exp(i\xiz_\rho(\bk)) \bb^\ad_\rho(-\bk)
\ea
\eeql
where
\beql{Col:th}
\ba{rcl@{\qquad}rcl}
\cos(2\th^y_\rho(\bk)) 
&=& \dsp {|k-\mu| \over \sqrt{ (k-\mu)^2 + F(k)^4 Q_\rho^2 }}, 
& \quad \xiy_\rho(\bk) &=& \phi(\bk)+\pi \\[1ex]
\cos(2\th^z_\rho(\bk)) &=& 
\dsp {k+\mu \over \sqrt{ (k+\mu)^2 + F(k)^4 Q_\rho^2 }},
& \quad \xiz_\rho(\bk) &=& -\phi(\bk) .
\ea
\eeql
These values are chosen so that $H$ has the form of a free Hamiltonian for
quasiparticles and quasiholes:
\beql{Col:quasiH}
\ba{rl}
H = \dsp\sum_{\bk,\rho}\Biggl\{ \phantom{+} &
    \sqrt{ (k-\mu)^2 + F(k)^4 Q_\rho^2} \,\, y^\ad_\rho(\bk)y_\rho(\bk) \\
+&  \sqrt{ (k+\mu)^2 + F(k)^4 Q_\rho^2} \,\, z^\ad_\rho(\bk)z_
\rho(\bk)\Biggr\}
\ea
\eeql
The true ground state $|\psi\>$ contains no quasiparticles:
\beql{Col:vac}
y_\rho(\bk) |\psi\> = z_\rho(\bk) |\psi\> = 0.
\eeql
The gap equations follow from requiring that
the mean field ansatz \eqn{Col:mf} hold in the quasiparticle basis.
In other words, we use \eqn{Col:yz} and \eqn{Col:th} to rewrite \eqn{Col:mf}
in terms of quasiparticle creation/annihilation operators, and then
evaluate the expectation value using \eqn{Col:vac}.
We get two gap equations,
\beql{Col:gap}
\ba{rcl}
 \De_8 + \quarter \De_1  &=& \dsp {4\over 3}K G(\De_1) \\[2ex]
 \eighth \De_1 &=& \dsp {4\over 3}K G(\De_8)
\ea
\eeql
where
\beq
G(\Delta) = -{1\over 2}\sum_{\bk}\Biggl\{
  {F(k)^2 \Delta\over \sqrt{ (k-\mu)^2 + F(k)^4 \Delta^2}}  
+ {F(k)^2 \Delta\over \sqrt{ (k+\mu)^2 + F(k)^4 \Delta^2} }\Biggr\}
\eeq
{}From \eqn{Col:quasiH} we see that the physical gap, namely the 
minimum energy of the quasiparticles, is $F(\mu)^2 \De$.
Creating a quasiparticle-quasihole pair requires at least
twice this energy.  The Nambu-Goldstone excitations 
discussed at length in Section~\ref{sec:ordering} are long
space-dependent oscillations of the phase and the flavor quantum
numbers of the condensate.  They do not involve
the excitation of quasiparticle-quasihole pairs.

\section{Results}
\label{sec:results}

We write our ``one-gluon'' coupling $K$ in terms of 
a dimensionless strong coupling constant $\al$:
\beq
K = {4\pi\al \over \La^2}
\eeq
If we think of this vertex as coming from integrating out the
gluons, then we expect $\al$ to be roughly of order 1.
As we will see below, the gap is sensitive 
to the coupling $\alpha$, and therefore to the 
choice of physical criterion used to fix $\alpha$.
The gap is also somewhat sensitive to details of the
form factor \eqn{ham:F}, as we shall see.  
Our methods therefore allow only an order of 
magnitude estimate of the magnitude of the gap.

We use the zero-density chiral symmetry breaking to fix the
coupling. If our single-gluon-exchange four-fermion vertex
\eqn{ham:H} is the only interaction in the theory, then
the chiral gap, which can play the role
of a constituent quark mass, is determined by the gap
equation
\beql{Ch:gap2}
{32 K\over 3} \sum_\bk {F(k)^2\over\sqrt{k^2 + F(k)^4\De_\chi^2}} = 1
\eeql
derived in Appendix A.
Since we are only keeping the part of the Hamiltonian that
models single-gluon exchange, we likely overestimate its strength if we
require that at zero density it give a phenomenologically reasonable 
value for
$\Delta_\chi$, say $0.4$ GeV. In reality, other
terms in the Hamiltonian
are at least partially responsible for the zero density chiral 
gap.
For example, although instanton effects are small at high density,
they certainly contribute to $\Delta_\chi$ at zero density. 
(For example, there is evidence\cite{appelquist} that in QCD with many flavors,
instantons contribute about as much as one-gluon exchange
to  $\Delta_\chi$ at zero density.)
We have therefore explored a range of couplings $\alpha$.

\begin{figure}[t]
\vspace{-0.75in}
\begin{center}
\epsfysize=4in
\epsfbox{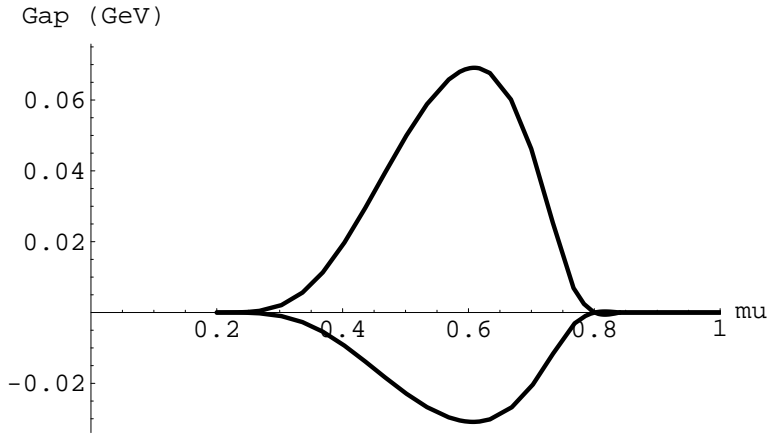}
\end{center}
\vspace{-15ex}
\caption{Physical color-flavor gaps as a function
of chemical potential, for a smoothed-step form factor  with 
scale $\La=0.8~\GeV$ 
and width $w=0.05~\GeV$.  The upper curve is $\Delta_1 F^2(\mu)$
while the lower curve is  $\Delta_8 F^2(\mu)$.
The coupling was chosen to be 
$\al = 0.252$, which is half of
the coupling at which our Hamiltonian would produce
a zero density chiral gap of 0.4~GeV. 
}
\label{fig:gaps_mu}
\end{figure}
\begin{figure}[thb]
\vspace{-0.75in}
\begin{center}
\epsfxsize=4in
\epsffile{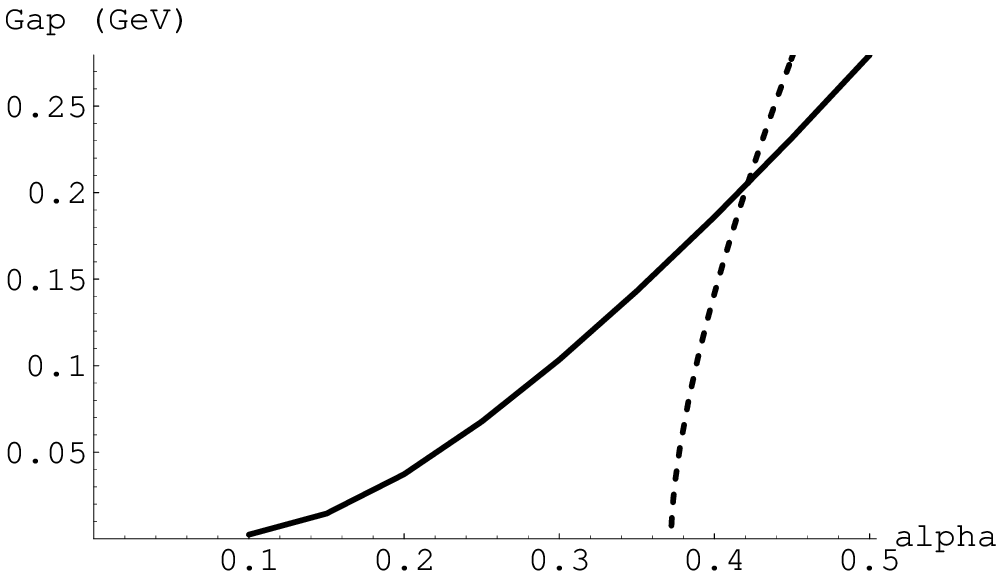}
\end{center}
\vspace{-15ex}
\caption{Color-flavor gap $\Delta_1 F(\mu)^2$ at
the $\mu$ at which it is largest (solid line)
and zero-density chiral gap (dashed line) 
as a function of coupling $\al$, for 
a smoothed-step form factor with scale $\La=0.8$ and 
width $w=0.05~\GeV$.}
\label{fig:maxgap}
\end{figure}

In Figure~\ref{fig:gaps_mu} we show how the color-flavor gaps vary with
chemical potential $\mu$ at fixed coupling: they are suppressed at low $\mu$ 
by the smallness of the Fermi surface, and at large $\mu$
by the fall-off of the form factor which
implements the weakening of the interaction 
that occurs in an asymptotically
free theory.  In comparing our results  
to those in the  two-flavor
theory, note that in Ref. \cite{ourselves} we plotted
the gap parameter $\Delta$, rather than the smaller, but physical,
gap $F^2(\mu)\Delta$.
In the two-flavor
theory with its single gap parameter,  the critical
temperature $T_c^\Delta$ above which the gap
vanishes is quite close to its BCS value 
of $0.57 F^2(\mu) \Delta(T=0)$ \cite{withjurgen}.
Here, it remains to be seen how $T_c^\Delta$, the
transition at which chiral and color symmetries are
restored, is related
to the two zero temperature gaps.
As in the two-flavor theory, we expect that at temperatures less
than $T_c^\Delta$, the 
chiral condensate is replaced by the superconducting
condensate above a first order phase transition which occurs at some $\mu_0$
which should be around $0.3$ GeV for a reasonable phenomenology
\cite{ourselves}.
However we have not included sufficient interactions to treat the low
density phase.  We therefore defer an estimate of the $\mu_0$ above
which the curves in Figure~1 are relevant.  Suffice to say that 
the chemical potentials at which the gaps reach their maximum
values are of physical interest.

Requiring $\alpha$ to be half that at which our Hamiltonian would
produce a zero density chiral gap of 0.4~GeV, as done in Figure~1, is
an ad hoc criterion.  In Figure~\ref{fig:maxgap} we therefore explore
the sensitivity of color-flavor condensation to the coupling, by
plotting the maximum of the gap.  The strongest possible coupling is
the one at which single-gluon-exchange alone gives the observed chiral
gap of about $0.4~\GeV$. Were $\alpha$ this large, the color-flavor
gap in the high density phase would be almost 300 MeV.   
We see that if we make the
single-gluon-exchange interaction weaker than that, 
by reducing $\al$ by a factor of two as in Figure~1,
the maximum of the gap is still significant, around
$70~\MeV$.  The gaps we find are comparable to those
found in the two-flavor theory with the 't~Hooft vertex being the only
interaction \cite{ourselves,instantonliquid,withjurgen}.  Single-gluon
exchange is of course also attractive in the two-flavor diquark
channel (\ref{oldCondensate}), which suggests that adding it to the
two flavor theory could enlarge the gap there too.

\def\st{\rule[-1.5ex]{0em}{4ex}} 
\begin{table}[t]
\begin{tabular}{ccccc}
\hline
\st $\Lambda$ & width $w$ & $\al$ & $\mu_{\rm max}$ & Max gap \\[1ex]
\hline
  0.8   &  0.05   &  0.252    &  0.605    &  0.0691 \\
  0.8   &  0.1   &  0.323    &  0.552    &  0.0512 \\
  0.8   &  0.2   &  0.490    &  0.503    &  0.0293 \\
  0.8   &  0.4   &  0.780   &  0.560    &  0.0154 \\
\hline
  0.5   &  0.05   &  0.363    &  0.357   &  0.0614 \\
  0.5   &  0.1   &   0.510    &  0.312    &  0.0398 \\
  0.5   &  0.2   &   0.794    &  0.336    &  0.0209 \\
  0.5   &  0.4   &   1.001     &  0.445    &  0.0119 \\
\hline
\st $\Lambda$ & $\nu$ & $\al$ & $\mu_{\rm max}$ & Max gap \\[1ex]
\hline
  0.8   &  0.75   &  0.364    &  0.581    &  0.0079 \\
  0.8   &   1.   &  0.580    &  0.472    &  0.0109 \\
  0.8   &   1.25   &  0.811    &  0.407    &  0.0131 \\
\hline
  0.5   &  0.75   &  0.409    &  0.362    &  0.0091 \\
  0.5   &   1.   &  0.668    &  0.298    &  0.0130 \\
  0.5   &   1.25   &  0.953    &  0.261    &  0.0160 \\
\hline
\end{tabular}
\vspace{3ex}
\caption{
Color-flavor gap ($F(\mu)^2 \De_1$), maximized with respect to $\mu$,
for a variety of form factors (see \eqn{ham:F}).
Top half: smoothed step; bottom half: power law.
In each case the coupling is chosen 
to be half of that
at which the chiral gap would be $0.4$ GeV at zero density.
All energies are in GeV.
}
\label{tab:robust}
\end{table}

In Table \ref{tab:robust} we vary the scale $\Lambda$
and the profile of the form factor, in each case
fixing the coupling $\al$ to be half that at which
the zero density chiral gap is 0.4~GeV.
We see that even once we use a physical
criterion to fix $\alpha$, the maximum gap 
as a function of chemical potential 
does depend on the details of the form factor, varying
by about a factor of nine for the form factors we have
considered.  Furthermore, we must consider the dependence
shown in Figure~2, namely the dependence on
the physical criterion by which $\alpha$ is fixed.
Were one to use an $\alpha$ such that the
zero density chiral gap in this model is comparable
to its physical value, gaps four to sixteen times larger
than those given in the Table would arise.  (The maximum
gaps
in the final column of the table would range from 
111 MeV to 284 MeV; note that with these larger
values of $\alpha$, the dependence on the shape
of the form factor is less severe.)
Our methods suggests that gaps of order ten to one hundred
MeV arise at accessible chemical potentials. To obtain
a more quantitative determination, one must either more firmly constrain
the couplings and
their form factors phenomenologically or, better, extend
the model to include the gluons themselves.

\section{Concluding Remarks}

In Section~\ref{sec:ordering} we have explained 
the consequences of the symmetry breaking
pattern which we propose for high density QCD.
All quarks have a gap. All gluons get a mass.
There is an octet of pseudoscalar approximate
Nambu-Goldstone bosons reflecting the
fact that chiral symmetry is broken because color and flavor
rotations are locked, and a singlet truly massless
scalar associated with the
breaking of baryon number.  There is a modified but still massless
photon; with respect to this photon, one finds only integrally charged
excitations.

We have seen that 
much remains to be done before a truly quantitative estimate
of the magnitude of the gaps is possible.  
In addition, there are qualitative reasons to add
the six-fermion 't~Hooft interaction, to add
a nonzero strange quark mass, to treat simultaneous
condensation in $\langle q C\gamma_5 q\rangle$ 
and $\langle \bar q q \rangle$ channels and to go beyond mean field
theory.  In order to study the long-wavelength physics
of this phase, we need to combine a
non-Abelian generalization of the Ginzburg-Landau
approach familiar from superconductivity with a suitably
modified chiral perturbation theory.  This could yield
insights into neutron star physics beyond the elementary
observations that direct neutrino emission is suppressed, the
hadronic material is transparent, and the thermal transport is
dominated by a neutral superfluid component.
A quantitative treatment relevant to heavy ion collisions
would also require the introduction of different
chemical potentials for each of the three flavors.
The system will then tend to evolve toward equal
chemical potentials, as in so doing it can
maximize the color-flavor gap and lower its energy.

The present qualitative considerations are already
striking.  The high density environment may be 
quite different than previously expected --- 
no massless gluons and no light quarks; physics
dominated by an octet of light pseudo-Nambu-Goldstone bosons and
a massless singlet.
In other words, QCD at high densities 
and low temperatures may in many ways be
much more similar to QCD at low densities than to
a weakly-coupled quark-gluon plasma.  If this is
the case, 
it may only be as heavy ion collisions reach higher
energies and create hotter and less dense 
conditions that they will be
able to access an approximately chirally symmetric phase with light
quark and gluon degrees of freedom.

\vspace{3ex}
\centerline{Acknowledgements}

We wish to thank J. Berges, R. L. Jaffe, and  A. Vainshtein for
helpful discussions.  We are also grateful to T. Sch\"afer
for noting errant factors of two in a previous version.

The research of MGA and FW
is supported in part by DOE grant DE-FG02-90ER40542;
MGA is also supported by the generosity of Frank and Peggy Taplin.
KR is supported in part by DOE cooperative
research agreement DE-FC02-94ER40818.

\appendix
\section{Chiral Gap equations at zero density}

As in Section~\ref{sec:SCgap}, use the Weyl-spinor form of the Hamiltonian,
although unlike in (\ref{HI})
the terms which contribute are now those in which 
gluon exchange couples a left-handed quark to a right-handed
quark.
Now make the mean-field ansatz that, for the true ground state $|\psi\>$,
\beql{Ch:mf}
\dsp {1\over V} \int \!d^3x\,
\<\psi| \psiLd^i_{\al\da}(x)\psiRd_j^{\de\da}(x) |\psi \>
= {3\over 32 K} \De_\chi \de^i_j \de^\de_\al\ , 
\eeql
where $\De_\chi$ will turn out to be the ordinary chiral gap
which can play the role of a constituent quark mass.
The interaction Hamiltonian becomes
\beq
H_I = \De_\chi \dsp \int \!d^3x\, \FF\, \psiRd_\be^{ib} \psiL_{ib}^\be + \cc
\eeq
Using \eqn{sp:modexp}, the Hamiltonian can be rewritten
\beql{Ch:H}
\ba{rcl}
 H &=& \dsp\sum_{\bk} k\Bigl( \aLd_{i\be}(\bk) \aL_{i\be}(\bk)
+ \bLd_{i\be}(\bk) \bL_{i\be}(\bk) \Bigr)
+ L\to R \\[3ex]
&-& \dsp \De_\chi \sum_{\bk}
F(k)^2\, \exp(-i\phi(\bk))\Bigl(
\aRd^i_\be(\bk) \bLd^\be_i(-\bk) + \bR^i_\be(\bk) \al^\be_i(-\bk)
\Bigr) + \cc,
\ea
\eeql
where the perturbative vacuum is annihilated by $\aL,\bL,\aR,\bR$.
Finally, we change basis to creation/annihilation operators 
$y$ and $z$ for quasiparticles and quasiholes, 
\beql{Ch:yz}
\ba{rcl}
\yL(\bk) &=& \cos(\th_L(\bk))\aL(\bk) 
- \sin(\th_L(\bk))\exp(i\xi_L(k))\bRd(-\bk) \qquad\hbox{and~~}L\to R\\[1ex]
\zL(\bk) &=& \sin(\th_L(\bk))\exp(i\xi_L(k))\aLd(\bk)
+ \cos(\th_L(\bk))\bR(-\bk) \qquad\hbox{and~~}L\to R
\ea
\eeql
where
\beql{Ch:th}
\ba{c}
\cos(2\th_L(\bk)) = \cos(2\th_R(\bk)) =
\dsp {k\over \sqrt{ k^2 + F(k)^4\De_\chi^2 }} \\[1.5ex]
\xi_L(\bk) = \phi(\bk),\qquad \xi_R(\bk) = -\phi(\bk)+\pi
\ea
\eeql
These values are chosen so that $H$ has the form of a free Hamiltonian for
quasiparticles and quasiholes of mass $\De_\chi$
\beq
H = \dsp\sum_{\bk,i,\be} \sqrt{k^2 + F(k)^4\De_\chi^2}
\Bigl(\yLd_{i\be}(\bk)\yL_{i\be}(\bk) 
+ \zLd_{i\be}(\bk)\zL_{i\be}(\bk) \Bigr) + L\to R
\eeq
The true vacuum $|\psi\>$ contains no quasiparticles:
\beql{Ch:ac}
\yL(\bk) |\psi\> = \zL(\bk) |\psi\> = 0, \qquad\hbox{and~~}L\to R
\eeql
The gap equation
\beql{Ch:gap}
{32 K\over 3} \sum_\bk {1\over\sqrt{k^2 + F(k)^4\De_\chi^2}} = 1
\eeql
follows.

\section{ Spinor conventions }

We use the spinor conventions of Ref. \cite{bailinbook}.
Using gamma matrices in the chiral representation,
\beq
\ba{rcl@{\quad}rcl@{\quad}rcl}
\ga^0 &=& \left( \ba{rr} 0 & I \\ I & 0 \ea \right) , &
\ga^i = - \ga_i &=& \left( \ba{rr} 0 & \si^i \\ -\si^i & 0 \ea \right), &
\ga_5 &=& \left( \ba{rr} -I & 0 \\ 0 & I \ea \right) ,
\ea
\eeq
quark fields are
\beq
\ba{rcl@{\qquad}rcl}
\psi &=& 
\left( \ba{c} \psi_{L\,a} \\[1ex] \psi_{R}^\da  \ea \right), &
\psi^\ad &=& 
\left( \ba{cc} \psi^\ad_{L\,\da} & \psi_{R}^{\ad\,a} \ea \right).
\ea
\eeq
For the 2-component Weyl spinors
$\psi_L$ and $\psi_R$, indices are raised and lowered by
\beq
\ep^{\da\dc} = \ep^{ac} =  
\left( \ba{rr} 0 & -1 \\ \phm 1 & 0 \ea \right),
\eeq
Expand the left-handed fermion field in modes, with color index $\al$ and
flavor index $i$:
\beql{sp:modexp}
\psi_{Li\al}(x) = {1\over \sqrt{V}} \sum_{\bk} 
\left(\ba{c}
 -\sin(\th/2)\,\e^{-i\phi} \\
 \cos(\th/2)
\ea\right)
 \left[
   a^{\phantom\ad}_{i\al}(\bk)  e^{-ik\cdot x}
+  b^\ad_{i\al}(\bk) e^{ik\cdot x}
\right]
\eeql
where $k_0 = |{\bf k}|$, and
$\th,\phi$ are the polar and azimuthal angles of $\bk$, so
under $\bk\to-\bk$, $\th \to \pi-\th$ and $\phi\to\pi+\phi$.
Here and throughout, $\sum_{\bk} \equiv \int d^3 k/(2\pi)^3$.

\end{document}